\documentclass[12pt,a4paper]{iopart}
\usepackage{graphicx}
\begin{document}

\title[Energy dependence of kaon production]{Energy dependence of kaon production in central Pb+Pb collisions}

\author{T~Kollegger for the NA49 Collaboration\footnote{For a full author list of the NA49 Collaboration see~\cite{AndreSQM}}}

\address{Institut f\"ur Kernphysik, August-Euler-Stra\ss e~6, 60486~Frankfurt, Germany}

\ead{kollegge@ikf.physik.uni-frankfurt.de}

\begin{abstract}
Recent results from the NA49 experiment on the energy dependence of charged kaon production in central Pb+Pb collisions are presented. First results from the new data at 80 $A\cdot$GeV beam energy are compared with those from lower and higher energies. A difference in the energy dependence of the $\left \langle K^{+} \right \rangle/\left \langle \pi^{+} \right \rangle$ and $\left \langle K^{-} \right \rangle/\left \langle \pi^{-} \right \rangle$ ratios is observed. The $\left \langle K^{+} \right \rangle /\left \langle \pi^{+} \right \rangle$ ratio shows a non-monotonic behaviour with a maximum near 40 $A\cdot$GeV. 
\end{abstract}



\section{Introduction}

The main aim of the heavy ion program at the CERN SPS is to find and characterise the transition from confined hadronic matter to a transient deconfined state of strongly interacting matter. 
When sufficiently high initial energy density is reached the for\-mation of a state of quasi free quarks and gluons, the quark gluon plasma (QGP), is expected. 
To search for signs of such a transition the NA49 experiment studies the hadronic observables of the collision for different beam energies (40, 80  and 158~A$\cdot$GeV) and different system sizes (p+p, p+A, C+C, Si+Si and Pb+Pb at various centralities).

Strangeness production has been proposed as one possible signature for such a transition~\cite{Raf82}. First experimental results from Pb+Pb (Au+Au) collisions at top SPS (158~A$\cdot$GeV) and top AGS (11~A$\cdot$GeV) energies have suggested that anomalies in strangeness production should take place between these energies \cite{GazI} which triggered the SPS energy scan program \cite{NA49Add1}. Within this ongoing project NA49 recorded central Pb+Pb collisions at 40 and 80~$A\cdot$GeV during the heavy ion runs in 1999 and 2000. The data at the top SPS energy (158~A$\cdot$GeV) was taken in previous runs. In this paper we report first results on the production of charged kaons in central Pb+Pb collisions at 80~A$\cdot$GeV and compare these new measurements with those from lower and higher energies.

\section{The NA49 large acceptance spectrometer}

\begin{figure}[bp]
\begin{center}
\includegraphics[width=\textwidth]{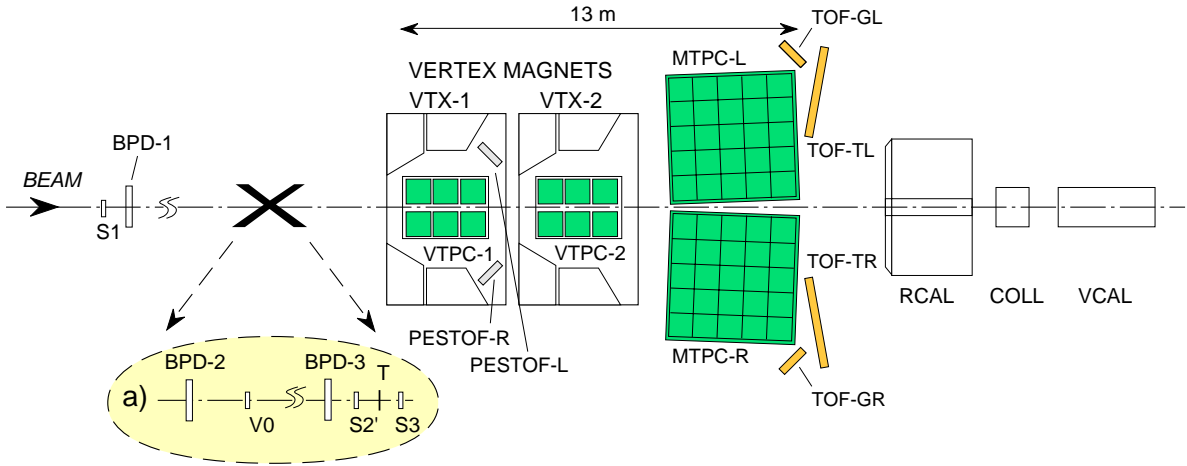}
\end{center}
\caption{\label{exp} Schematic diagram of the NA49 experimental apparatus in the configuration used for heavy ion runs. The main tracking devices are four large-volume time projection chambers VTPC-1, VTPC-2, MTPC-L and MTPC-R. Particle identification capabilities are augmented by the time of flight (TOF) detectors (see text for further details).}
\end{figure}

The NA49 experimental set--up \cite{NA49NIM} (see figure~\ref{exp}) consists of four large volume Time Projection Chambers (TPCs). Two of them, Vertex TPCs (VTPC-1 and VTPC-2), are placed in the magnetic field of two super-conducting dipole  magnets and allow a precise measurement of particle momenta $p$ ($\sigma(p)/p^2 \approx (0.3-7)\cdot10^{-4}$ (GeV/c)$^{-1}$) and electric charge.  The other two TPCs (MTPC-L and MTPC-R), positioned downstream of the magnets, were optimised for high precision detection of ionization energy loss $dE/dx$ (relative resolution of about 4\% ) and consequently provide a means of measuring the particle  mass. The TPC data yield spectra of identified hadrons above midrapidity. The particle identification capability of the MTPCs is augmented by two Time of Flight (TOF) detector arrays (resolution $\sigma_{TOF} \approx $ 60 ps). The combined TPC and TOF information allow for the measurement of charged kaon spectra at midrapidity. 

In order to optimise the NA49 acceptance at 40 and 80 A$\cdot$GeV the magnetic fields in the VTPCs were lowered in proportion to the beam energy. The configuration at 158 A$\cdot$GeV was: B(VTPC-1) $\approx$ 1.5 T and B(VTPC-2) $\approx$ 1.1 T. The target, a thin lead foil (224 mg/cm$^2$), was positioned about
80 cm upstream from VTPC-1. Central collisions were selected by a trigger using information from a downstream calorimeter (VCAL), which measured the energy 
of the projectile spectator nucleons. The geometrical acceptance of the Veto Calorimeter was adjusted for each energy by a proper setting of a collimator (COLL). The datasets used in this analysis are the most 7\% (7\%, 5\%) central collisions of all inelastic interactions at 40 (80, 158) A$\cdot$GeV.

\section{Kaon spectra}

Raw $K^+$ and  $K^-$ yields were extracted from  fits of the 
distributions of $\textrm{d}E/\textrm{d}x$ and TOF (where available) in narrow 
bins of rapidity $y$ and transverse momentum $p_{{T}}$. 
The resulting spectra were then
corrected for geometrical acceptance, losses due to in--flight decays,
reconstruction efficiency ($\cong$ 5\%) \cite{TKDipl}.

\begin{figure}[htbp]
\begin{center}
\includegraphics[width=\textwidth]{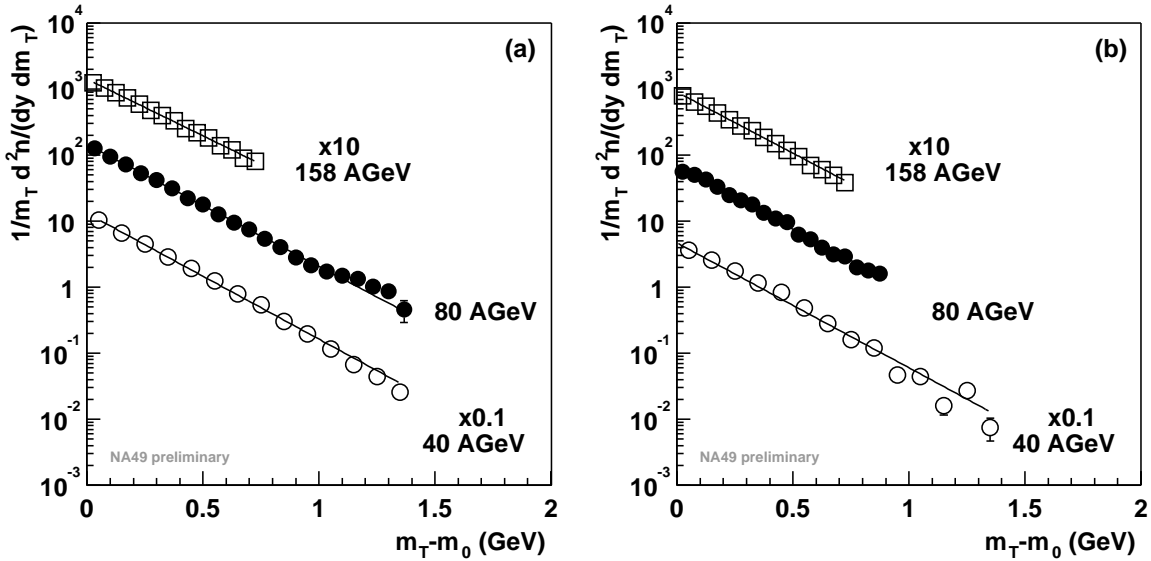}
\end{center}
\caption{\label{kmt} Transverse mass spectra at midrapidity for $K^{+}$ (a) and $K^{-}$ (b) mesons. Together with the results at 80~A$\cdot$GeV (\fullcircle) the scaled results from 40 (\opencircle, $\times 0.1$)\cite{NA49QM01} and 158~A$\cdot$GeV (\opensquare, $\times 10$)\cite{NA49QM99} are plotted. The solid line indicates the fit results (see text for further details)}
\end{figure}

The transverse mass ($m_{{T}} = \sqrt{p_{{T}}^{2}+m_{0}^{2}}$, $m_{0}$ is the rest mass of the particle) spectra of $K^{+}$ and $K^{-}$ mesons produced at midrapidity ($\mid y^{*} \mid < 0.1$) in central Pb+Pb collisions at 80 A$\cdot$GeV are shown in \fref{kmt}. The solid line indicates the result of a fit of the function $1/m_{{T}} \cdot d^{2}n/(dydm_{{T}}) = C \cdot \exp(-m_{{T}}/T)$. Also shown are the results of the NA49 measurements at 40 and 158~A$\cdot$GeV \cite{NA49QM01,NA49QM99} fitted with the same function. The obtained values of the inverse slope parameter $T$ range between 220 and 240 MeV.

\begin{figure}
\begin{center}
\includegraphics[width=\textwidth]{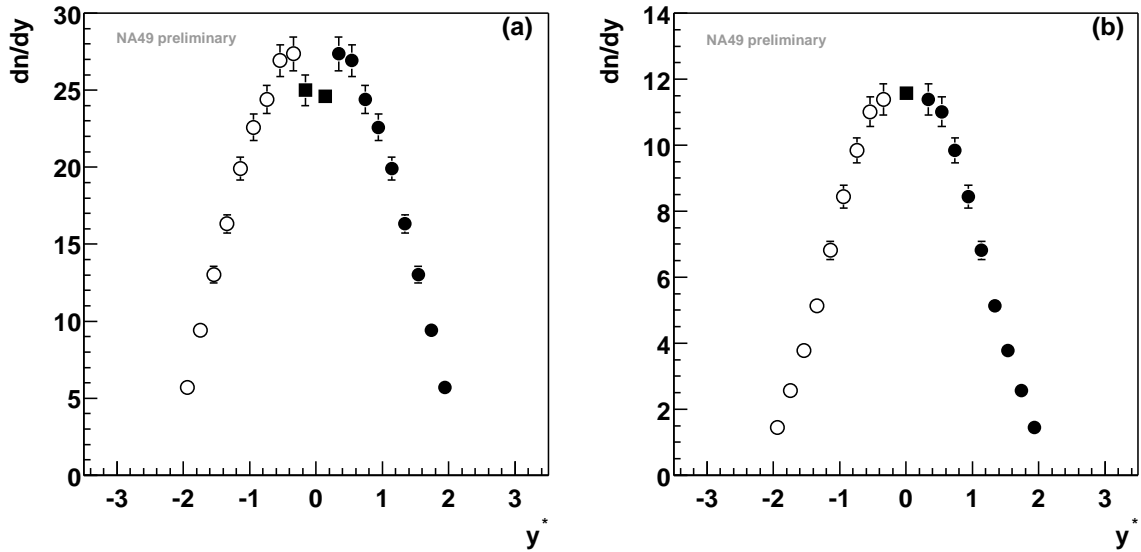}
\end{center}
\caption{\label{krap} Rapidity spectra of $K^{+}$ (a) and $K^{-}$ (b) mesons produced in 80~$A\cdot$GeV Pb+Pb collisions. Full circles (\fullcircle) show the results of the $\textrm{d}E/\textrm{d}x$-only analysis, squares (\fullsquare) the results from combined TOF-$\textrm{d}E/\textrm{d}x$ analysis. The measured $y$ spectra are reflected at midrapidity, shown by the open symbols (\opencircle)}
\end{figure}

The rapidity distributions $\textrm{d}n/\textrm{d}y$ were obtained by integrating the measured $m_{T}$ spectra in different rapidity bins. \Fref{krap} shows the rapidity distribution of $K^{+}$ and $K^{-}$ mesons produced  in central Pb+Pb collisions at 80 A$\cdot$GeV. A small systematic difference between the $\textrm{d}E/\textrm{d}x$-only and the combined TOF-$\textrm{d}E/\textrm{d}x$ measurements is visible, especially for the $K^{+}$ which needs further investigation. The mean kaon multiplicities in full phase space were derived by integration of the measured rapidity spectra. A necessary correction to the part of the spectrum not covered by the NA49 acceptance was applied based on a fit of two Gaussian functions to the rapidity distributions. The extracted total yields are $\left \langle K^{+} \right \rangle = 79 \pm 5$ and $\left \langle K^{-} \right \rangle = 29 \pm 2$.

\section{Pion spectra}

Due to limitations of the $\textrm{d}E/\textrm{d}x$ method
a somewhat different procedure was used to obtain
spectra of $\pi^-$-mesons. Yields of all negatively charged particles 
were determined as a function of $y$ (assuming the $\pi$ mass) and $p_T$
and corrected for geometrical acceptance.
The contamination of vertex $K^-$, $\overline{p}$
and $e^-$ as well as non-vertex hadrons originating from strange
particle decays and secondary interactions was subtracted.
The final spectra are also corrected for 
tracking and quality cut losses ($\cong$5\%) \cite{RolliDipl}.

\begin{figure}
\begin{center}
\includegraphics[height=8cm]{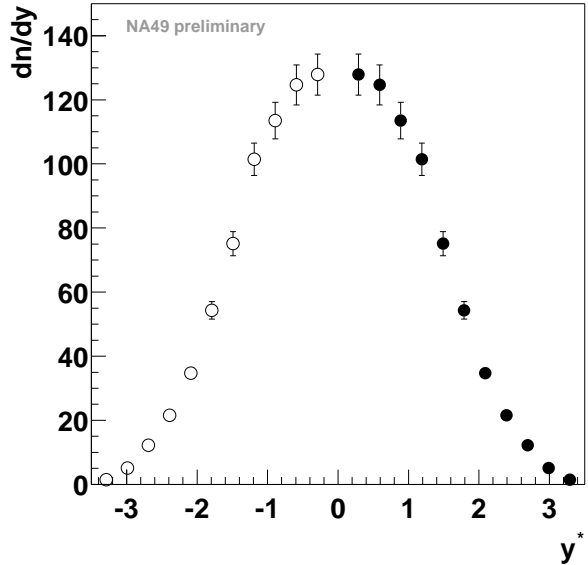}
\end{center}
\caption{\label{pi} Rapidity spectra of $\pi^{-}$ mesons produced in 80~$A\cdot$GeV Pb+Pb collisions. The measured $y$ spectra (\fullcircle) are reflected at midrapidity, shown by open symbols (\opencircle). }
\end{figure}

The rapidity distribution of negatively charged pions in central Pb+Pb collisions at 80 A$\cdot$GeV is shown in \fref{pi}. The total yield is $\left \langle \pi^{-} \right \rangle = 445 \pm 22$. The $\left \langle \pi^{+} \right \rangle / \left \langle \pi^{-} \right \rangle$ ratio could be determined in phase space regions where TOF-$\textrm{d}E/\textrm{d}x$ identification works (forward rapidity region starting slightly above midrapidity). In 80~A$\cdot$GeV central collisions it is $0.93 \pm 0.04$.

\section{Energy dependece of kaon production}

\begin{figure}[h]
\begin{center}
\includegraphics[width=\textwidth]{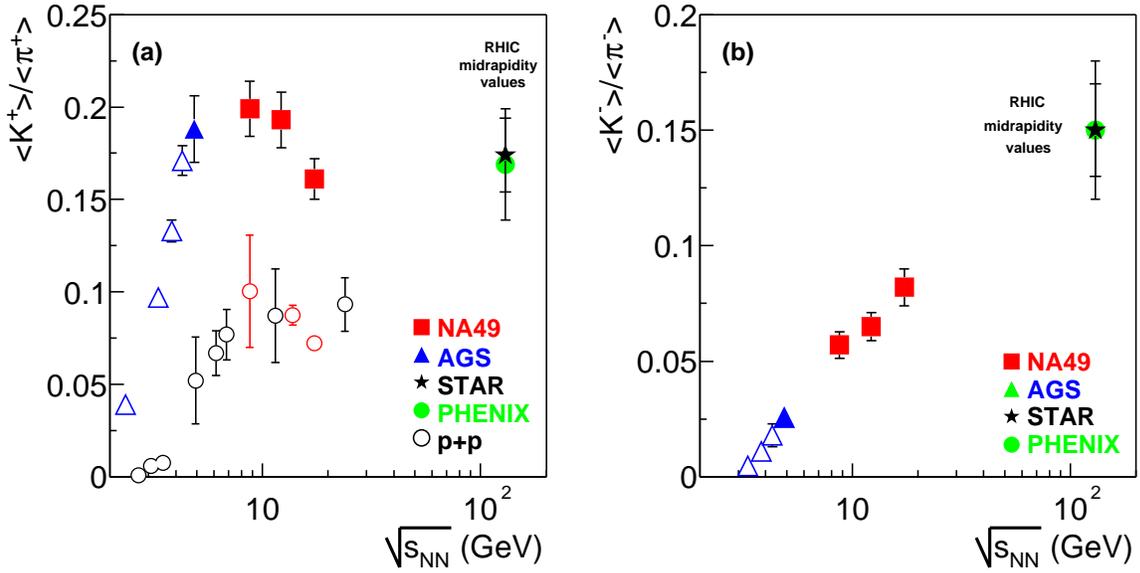}
\end{center}
\caption{\label{kpi} Energy dependence of the $\left \langle K^{+}\right \rangle/\left \langle \pi^{+}\right \rangle$ (a) and the $\left \langle K^{-}\right \rangle/\left \langle \pi^{-}\right \rangle$ (b) ratios. Integrated values were used for the ratios, except for RHIC where only midrapidity values are available.}
\end{figure}

\begin{figure}[h]
\begin{center}
\includegraphics[width=\textwidth]{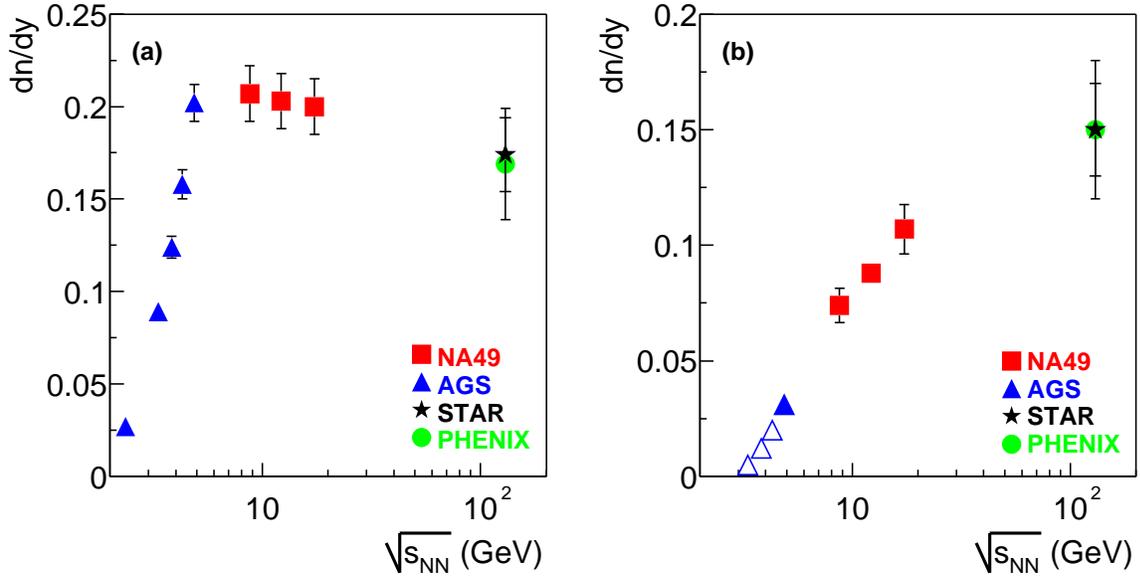}
\end{center}
\caption{\label{kpi_mid} Energy dependence of the $K^{+}/\pi^{+}$ (a) and the $K^{-}/\pi^{-}$ (b) ratios at midrapidity.}
\end{figure}

\Fref{kpi} shows the energy dependence of the $\left \langle K \right \rangle/\left \langle \pi \right \rangle$ full phase space ratio both for positively charged particles (a) and negatively charged ones (b). The data for the lower energies are taken from various AGS experiments \cite{AGS1} and extrapolated to full phase space where not measured (open symbols). For the higher energies only preliminary results at midrapidity from STAR \cite{STAR1} and PHENIX \cite{Phenix} at RHIC are published. They are nevertheless shown in the figure for comparison. For the $\left \langle K^{+}\right \rangle /\left \langle\pi^{+}\right \rangle $ ratio results \cite{NA49QM01} from p+p collisions at different energies are also shown. The energy dependence of the ratios at midrapidity is shown in \fref{kpi_mid}.

A different behaviour of the $\left \langle K^{+}\right \rangle /\left \langle \pi^{+}\right \rangle$ and the $\left \langle K^{-}\right \rangle/\left \langle \pi^{-}\right \rangle$ ratio is observed. The $\left \langle K^{+}\right \rangle/\left \langle \pi^{+}\right \rangle$ ratio shows a non-monotonic behaviour with a steep increase in the low energy (AGS) region followed by a turnover (around 40~$A\cdot$GeV or $\sqrt{s_{NN}}=8.73$ GeV) into a decreasing trend. The $\left \langle K^{-}\right \rangle/\left \langle \pi^{-}\right \rangle$ ratio is monotonically increasing. A similar behaviour is observed for the midrapidity ratios.

\begin{figure}
\begin{center}
\includegraphics[height=8cm]{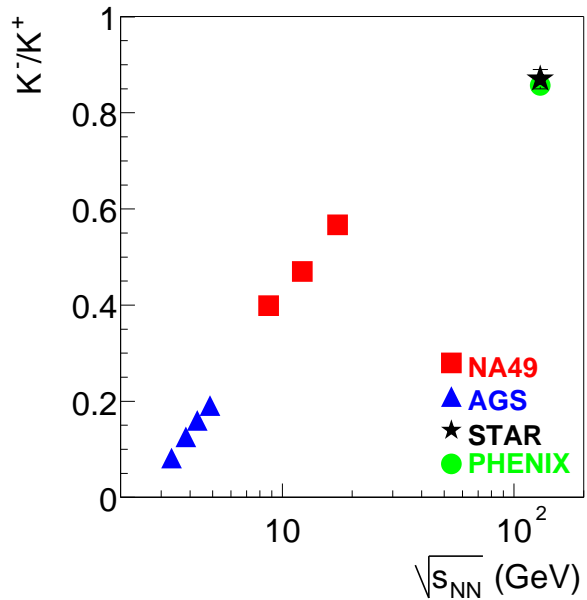}
\end{center}
\caption{\label{kmkp_mid} Energy dependence of the $K^{+}/K^{-}$ ratio at midrapidity.}
\end{figure}

This different energy dependence of the $\left \langle K^{+}\right \rangle/\left \langle \pi^{+}\right \rangle$ and $\left \langle K^{-}\right \rangle/\left \langle \pi^{-}\right \rangle$ ratios could be attributed to their different sensitivity on baryon density. Kaons ($K^{+}$ and $K^{0}$) carry more than 80\% (or more than 90 \% if only open strangeness is considered) of all produced $\bar s$-quarks. The $K^{+}$ yield is therefore nearly proportional to the total strangeness production. This picture changes if one looks at $s$-quarks, where only a much smaller fraction is carried by antikaons ($K^{-}$ and $\bar K^{0}$) whereas hyperons carry a significant fraction. If one assumes strangeness conservation (i.e. the same amount of $s$ and $\bar s$ quarks), the decreasing trend in the $\left \langle K^{+}\right \rangle/\left \langle \pi^{+}\right \rangle$ ratio above 40~$A\cdot$GeV and the increase in the $\left \langle K^{-}\right \rangle/\left \langle \pi^{-}\right \rangle$ ratio should result in a decreasing hyperon to pion ratio. The most abundant hyperons are lambdas and preliminary results from NA49 \cite{AndreSQM,KresoSQM} indeed show the expected decrease. The different sensitivity on baryon density should also result in an increase of the $\left \langle K^{-} \right \rangle / \left \langle K^{+} \right \rangle$ ratio. \Fref{kmkp_mid} shows this ratio at midrapidity and indeed an increase is observed.

Various models have been used to describe the energy dependence of strangeness production with and without invoking a transient deconfined or QGP phase. The extended statistical hadron gas model \cite{KPiTherm} captures the trend of the data, however the decrease in the full phase space ratio between 40 and 158~$A\cdot$GeV is not well reproduced. The measured $\left \langle K^{+}\right \rangle/\left \langle \pi^{+}\right \rangle$ ratio at 158~$A\cdot$GeV is 25\% lower than predicted by the model. The RQMD model \cite{RQMD,KPiRQMD} predicts a monotonic increase of the $\left \langle K^{+}\right \rangle/\left \langle \pi^{+}\right \rangle$ ratio between top AGS and top SPS energies and overpredicts the measurement at 158~$A\cdot$GeV by about 25\%. UrQMD \cite{UrQMD,KPiUrQMD} overestimates pion production and predicts therefore a too small $\left \langle K^{+}\right \rangle/\left \langle \pi^{+}\right \rangle$ ratio (e.g. by about 40\% too low at 40 $A\cdot$GeV) and shows no significant energy dependence above top AGS energy. From the HSD model \cite{KPiHSD}, only predictions for the midrapidity $\left \langle K^{+}\right \rangle/\left \langle \pi^{+}\right \rangle$ ratio are available, it shows a monotonic increase with energy. The schematic Statistical Model of the Early Stage \cite{Marek} assumes that a transition from a reaction with purely confined matter to a reaction with a QGP at the early stage occurs close to 40 A$\cdot$GeV. It predicts a rapid change of the strangeness to entropy ratio (closely proportional to the $\left \langle K^{+}\right \rangle/\left \langle \pi^{+}\right \rangle$ ratio) near the transition energy and a constant value once the threshold for deconfinement has been crossed.

\section{Summary and conclusions}

We have presented first results on charged kaon and pion production in central Pb+Pb collisions at 80~A$\cdot$GeV. The total yields are $\left \langle K^{+} \right \rangle=79 \pm 5$, $\left \langle K^{-} \right \rangle=29 \pm 2$ and $\left \langle \pi^{-} \right \rangle=445 \pm 22$. The observed energy dependence of the $\left \langle K^{+}\right \rangle/\left \langle \pi^{+}\right \rangle$ ratio shows a non-monotonic behaviour with a maximum around 40~A$\cdot$GeV followed by a decreasing trend. The $\left \langle K^{-}\right \rangle/\left \langle \pi^{-}\right \rangle$ ratio shows a monotonic increase with collision energy. Such a difference is expected due to the different sensitivity on net baryon density. NA49 will continue to study the energy dependence in 2002 with runs at 20 and 30~A$\cdot$GeV.

\ack
This work was supported by the Director, Office of Energy Research, 
Division of Nuclear Physics of the Office of High Energy and Nuclear Physics 
of the US Department of Energy (DE-ACO3-76SFOOO98 and DE-FG02-91ER40609), 
the US National Science Foundation, 
the Bundesministerium f\"ur Bildung und Forschung, Germany, 
the Alexander von Humboldt Foundation, 
the UK Engineering and Physical Sciences Research Council, 
the Polish State Committee for Scientific Research (5 P03B 13820 and 2 P03B 02418), 
the Hungarian Scientific Research Foundation (T14920 and T23790),
the EC Marie Curie Foundation,
and the Polish-German Foundation.

\end{document}